\documentclass[aps,prx,twocolumn,groupedaddress,showkeys,showpacs]{revtex4-1}
\usepackage{graphicx}
\usepackage{xcolor}
\usepackage{dcolumn}
\usepackage{graphics}
\usepackage{epstopdf}
\usepackage{amssymb}
\usepackage{amsmath}
\usepackage{bm}
\usepackage{float}
\usepackage{braket}
\usepackage{marvosym }

\begin{document}
\title{Bound states in the continuum and Fano resonances \\ in the strong mode coupling regime}

\author{Andrey~A.~Bogdanov${}^{1,2}$}
%\email{a.bogdanov@metalab.ifmo.ru}
\author{Kirill~L.~Koshelev${}^{1,3}$}
\author{Polina~V.~Kapitanova${}^{1}$}
\author{Mikhail~V.~Rybin${}^{1,2}$}
\author{Sergey~A.~Gladyshev${}^{1}$}
\author{Zarina~F.~Sadrieva${}^{1}$}
\author{Kirill~B.~Samusev${}^{1,2}$}
\author{Yuri~S.~Kivshar${}^{1,3}$}
\email{\textcolor{black}{ysk@internode.on.net}}
\author{Mikhail~F.~Limonov${}^{1,2}$}

\affiliation{$^1$ITMO University, St.Petersburg 197101, Russia}
\affiliation{$^2$Ioffe Institute, St.~Petersburg 194021, Russia}
\affiliation{$^3$Nonlinear Physics Center, Australian National University, Canberra ACT 2601, Australia}

%\date{\today}
\begin{abstract}
The study of resonant dielectric nanostructures with high refractive index  is a new research direction in  nanoscale optics and metamaterial-inspired nanophotonics. Because of the unique optically-induced electric and magnetic Mie resonances, high-index nanoscale structures are expected to complement or even replace different plasmonic components in a range of potential applications.  Here we study strong coupling  between modes of a single subwavelength high-index dielectric resonator and analyse the mode transformation and Fano resonances when resonator's aspect ratio varies. We demonstrate that  strong mode coupling results in resonances with high quality factors, which are related to the physics of \textit{bound states in the continuum} when the radiative losses are almost suppressed due to  the Friedrich--Wintgen scenario of destructive interference. \textcolor{black}{We explain the physics of these states in terms of multipole decomposition and show that their appearance is accompanied by drastic change of the far-field radiation pattern.} We reveal a fundamental link between the formation of the high-quality resonances and peculiarities of the Fano parameter in the scattering cross-section spectra. Our theoretical findings are confirmed by microwave experiments for the scattering of a high-index cylindrical resonators with a tunable aspect ratio. The proposed mechanism of the strong mode coupling in single subwavelength high-index resonators accompanied by resonances with high quality factor helps to extend substantially functionalities of all-dielectric nanophotonics that opens new horizons for active and passive nanoscale metadevices.
\end{abstract}
%\pacs{42.25.Fx,41.20.Jb,42.79.Dj}
 \maketitle
%%%%%%%%%%%%%%%%%%%%%%%%%%%%%%%%%%%%%%%%%%%%%%%%%%%%%%%%%%%%%%%%%%%%%%%%%%%%%%%%%%%%%%%%%%%%%%%%%%%%%%
\section{Introduction}
The physics of resonant structures with strong mode coupling is of the fundamental importance being responsible for a variety of interesting phenomena governing both transport and localization of waves. The modes supported by traditional resonators and microcavities~\cite{Vahala2003} exist due to reflection of waves from the resonator's boundaries under the conditions of constructive interference. To achieve high values of resonator's quality factor ($Q$ factor), one can improve reflectivity by using metals~\cite{Min2009,kwon2010subwavelength}, photonic bandgap structures~\cite{akahane2003high}, or the total internal reflection at glancing angles of incidence in whispering-gallery-mode (WGM) resonators~\cite{Matsko2006}. All those physical mechanisms require large sizes of cavities with a complex design. A more attractive way to confine light is to use destructive interference in the regime of strong mode coupling~\cite{Lai1991,Johnson2001, Rybin2017}. This mechanism is related to the physics of {\it bound states in the continuum} (BIC)~\cite{von1929uber}. It was first proposed in quantum mechanics by Friedrich and Wintgen~\cite{Friedrich1985} and then was extended to acoustics~\cite{parker1966resonance,parker1967resonance,Lyapina2015} and electrodynamics~\cite{Marinica2008,Bulgakov2008}.  A true optical BIC is a mathematical abstraction since its realization demands either infinite size of the structure or zero (or infinite) permittivity~\cite{Ndangali2010, Hsu2013, Monticone2014}. However, the BIC-inspired mechanism of light localization makes possible realization of high-$Q$ states in photonic crystal cavities and slabs~\cite{Bulgakov2008,Hsu2013,rybin2017optical}, coupled waveguide arrays~\cite{Plotnik2011,molina2012surface,corrielli2013observation}, dielectric gratings~\cite{Marinica2008}, core-shell spherical particles~\cite{Monticone2014}, dielectric resonators~\cite{wiersig2006formation, unterhinninghofen2008goos, Lepetit2014, lepetit2010resonance}, \textcolor{black}{and hybrid plasmonic-photonic systems~\cite{plasmonics}}.

By definition, the $Q$ factor of a true BIC is infinite. Hence, in the wave scattering, BIC manifests itself as ``a collapse'' of the Fano resonance when the width of the resonance vanishes and the Fano feature disappears from the scattering spectrum~\cite{Fonda1963,kim1999cs}.  { In practice, both Q factor and the width of the Fano resonance at the frequency of BIC remain finite because of absorption, finite-size samples, roughness, and other imperfections~\cite{Sadrieva2017}.} Remarkably, in terms of the scattering matrix, BIC corresponds to merging of a pole and a zero of the scattering operator on the real axis~\cite{Blanchard2016}. Properties of the Fano resonances in the systems with BICs have been considered in several studies~\cite{Bulgakov2008,Hsu2013,Monticone2014}. 

\begin{figure*}[t]
   \centering
   \includegraphics[width=0.7\linewidth]{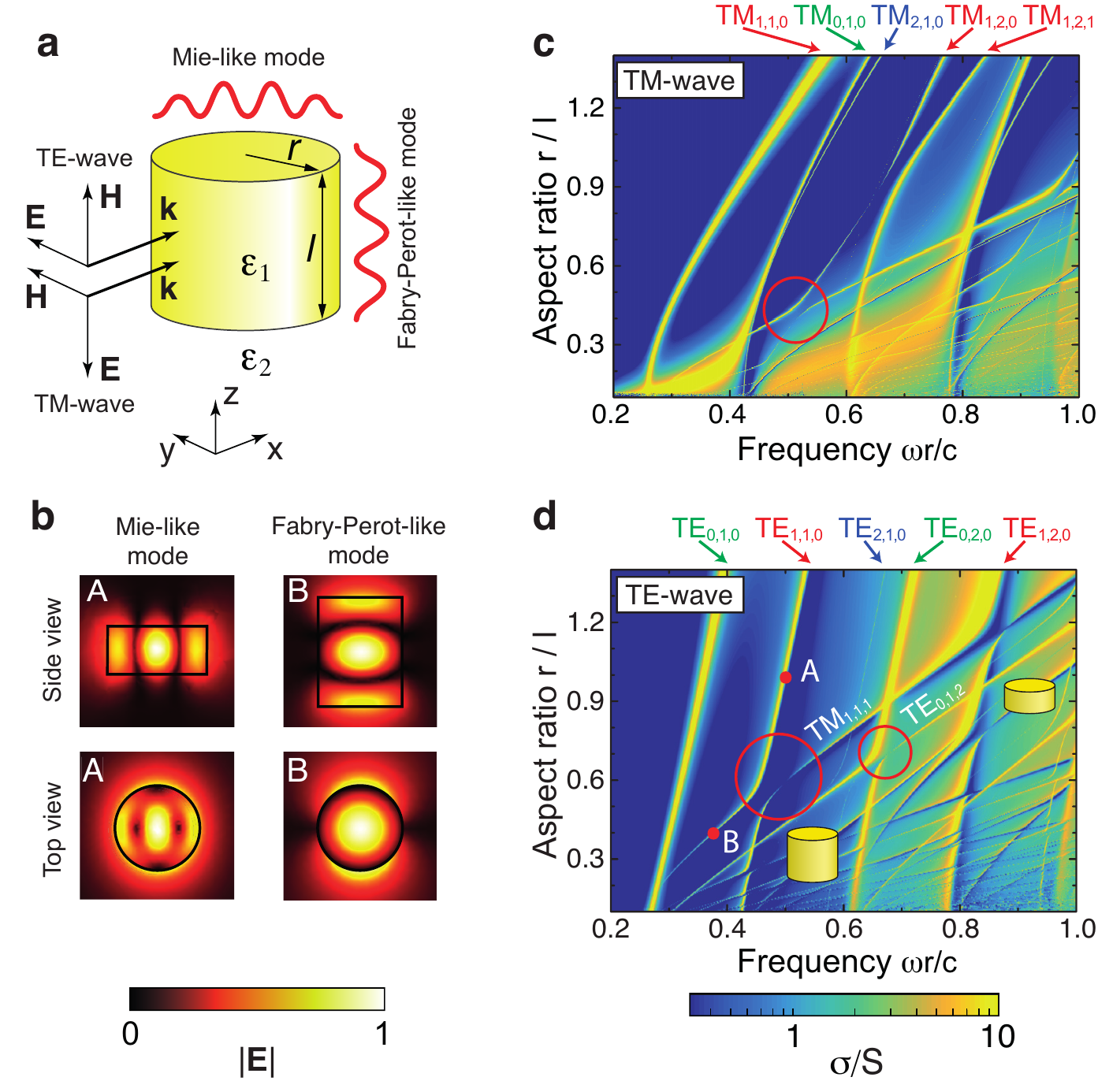} % requires the graphicx package
   \caption{ {Strong coupling of modes in a dielectric resonator.} ({a}) TE- and  TM-polarized waves incident on a dielectric cylindrical resonator with permittivity $\varepsilon_1=80$, radius $r$, and length $l$ placed in vacuum ($\varepsilon_2=1$). ({b}) Distribution of the electric field amplitude $|\mathbf{E}|$ for the Mie-like mode TE$_{1,1,0}$ (point A) and Fabry-Perot-like mode TM$_{1,1,1}$  (point B). ({c, d}) Dependencies of the total scattering cross-section (SCS) of the cylinder $\sigma$ normalized to the projected cross-section $S=2rl$ on the aspect ratio of the cylinder and frequency $r\omega/c=2\pi r/\lambda$ for TM and TE-polarized incident wave, respectively. The calculations are carried out with the step of $r/l=0.003$.  In panels ({c, d}), the regions of the most pronounced avoided crossing are marked by red circles.}
   \label{fig:figure_1}
\end{figure*}

 A conventional device supporting light localization via BIC-inspired mechanism is based on periodic photonic structures~\cite{Hsu2013} or chains of scatterers~\cite{bulgakov2017propagating}. For these structures, strong localization can be achieved only for a large number of scatterers because it is governed by their mutual interference. Other implementations  of BICs in photonic structures were presented in Refs.~\cite{Lepetit2014,Monticone2014}. In the former work, the design is based on  a metallic box with a single dielectric scatterer inside. While the conditions of true BICs can be achieved here by imitating periodic boundary conditions for such a unit cell, the structure is not subwavelength, and it demands using metallic components. The latter work describes a single scatterer but it employs near-zero refraction index constituents which requires large scales of the structure~\cite{liberal2017near}, especially, at optical frequencies.  Recently BIC-inspired supercavity modes in subwavelength dielectric resonators without singular permittivity values were proposed theoretically \cite{Rybin2017}. However, the coexisting of Fano resonance being effect of weak coupling and strong mode coupling underlying BIC is a vital problem of modern photonics.

In this paper we demonstrate, both theoretically and experimentally,  {strong light localization} and existence of quasi-BICs in the simplest object -- a single homogeneous cylindrical subwavelength dielectric resonator in free space. We show that quasi-BICs appear in accord with the Friedrich--Wintgen interference mechanism because of strong coupling between Mie-like and Fabry-Perot-like modes.   We develop a novel analytical approach to describe light scattering by finite-size dielectric resonators and reveal close relationships between the peculiarities of quasi-BICs and the critical behaviour of Fano asymmetry parameter in the strong coupling regime. We show that the Fano asymmetry parameter becomes singular at the frequency of the quasi-BIC, and it vanishes when the mode becomes almost dark for far field excitation. We derive an exact form of coupling coefficients between modes and corresponding Rabi frequencies. We analyse effects of material losses and reveal that the mode coupling remains strong even for highly absorptive resonators. Our findings make evident the counterintuitive fact that even a subwavelength dielectric resonator could provide strong light localization.

\begin{figure}
   \centering
   \includegraphics[width=1\linewidth]{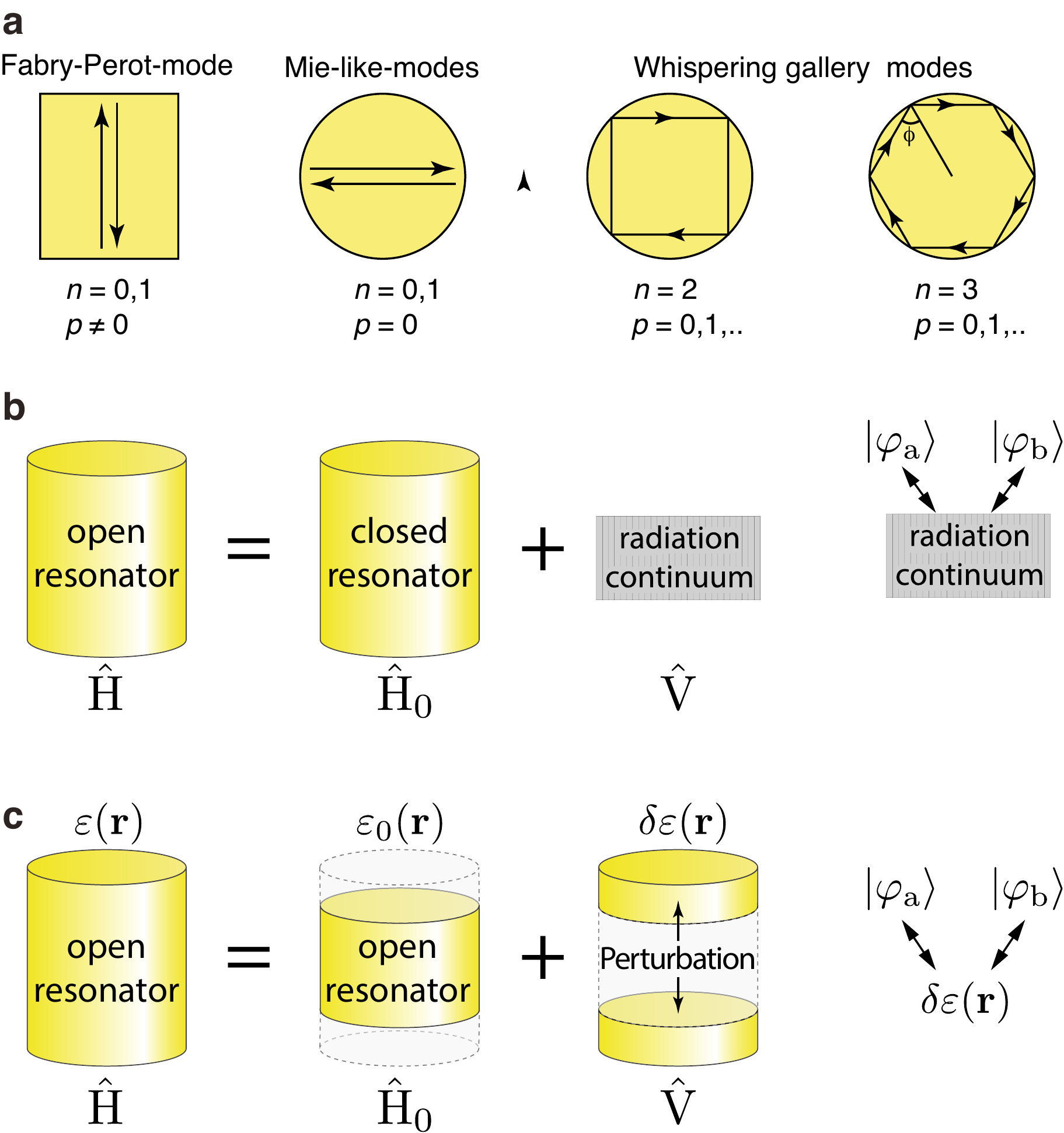} % requires the graphicx package
   \caption{ {Modes of a dielectric resonator and models of their coupling}. ({a}) Classification of eigenmodes of a dielectric resonator.  ({b}) Friedrich--Wintgen approach describing an open cylindrical resonator as a closed resonator and a radiation continuum. Eigenmodes of the resonator interact via the radiation continuum. ({c}) Non-Hermitian approach describing an open cylindrical resonator by a complex spectrum of eigenfrequencies. Eigenmodes of the resonator interact via perturbation $\delta\varepsilon(\mathbf{r})$ responsible for change of resonator aspect ratio.}
   \label{fig:figure_11}
\end{figure}

%==================================
\section{Results}
\subsection{Interplay of Mie and Fabry-Perot modes}

We start our study with numerical simulations of the scattering cross-section (SCS) $\sigma$ of a finite dielectric cylinder with permittivity $\varepsilon_1=80$, radius $r$, and length $l$ placed in vacuum ($\varepsilon_2=1$), as shown in Fig.~\ref{fig:figure_1}a. The spectra are calculated by using the CST Microwave Studio software and T-matrix computations~\cite{mishchenko1993light,mishchenko1994t}. The electric field of the incident wave is assumed to be perpendicular to the axis of the cylinder (see Fig.~\ref{fig:figure_1}a). To compare $\sigma$ for cylinders with different aspect ratios, we normalize $\sigma$  by the projected cross-section of the resonator, $S=2rl$.  The maps of the normalized SCS $\sigma/S$ calculated for cylinders with different aspect ratio $r/l$ excited by TM and TE-polarized wave are shown in Figs.~\ref{fig:figure_1}c and \ref{fig:figure_1}d, respectively. According to the standard nomenclature (see, e.g., Ref.~\cite{zhang2008electromagnetic}), we denote the modes of a cylindrical resonator as TE$_{n,k,p}$ and TM$_{n,k,p}$, where $n, k, p$ are the indices denoting the azimuthal, radial, and axial wavenumbers, respectively. Generally speaking, distinguishing between TE$_{n,k,p}$ and TM$_{n,k,p}$ modes for a cylinder of a finite length is justified only for $n=0$. For other cases, the polarisation is hybrid~\cite{jackson2007classical}. In the case of arbitrary $n,k,p$ the mode polarization is mixed. Thus, under the terms ``TE'' or ``TM'' we further imply the dominant polarization of the modes.

\begin{figure*}
   \centering
  \includegraphics[width=0.8\linewidth]{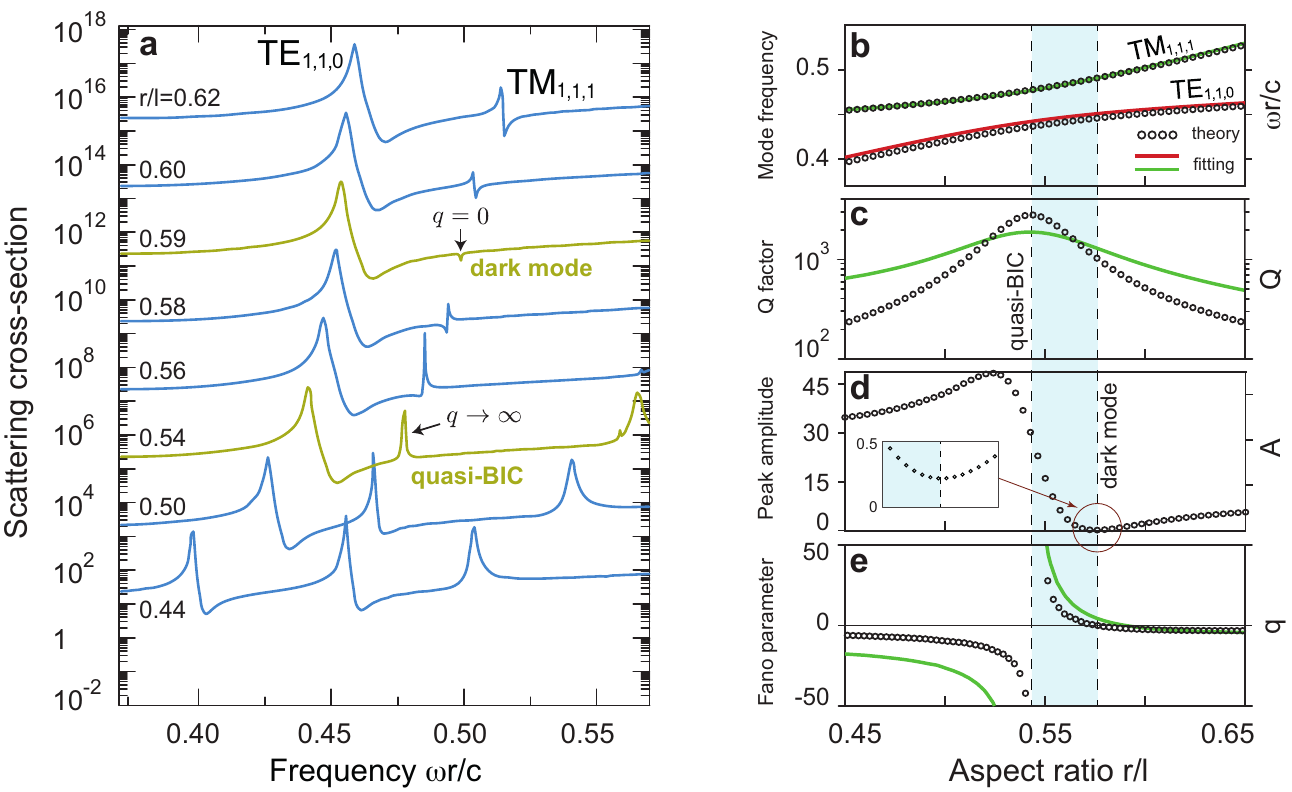} % requires the graphicx package
   \caption{ {Avoided resonance crossing, $\mathbf{Q}$-factor and Fano resonance}. ({a}) Spectra of the normalized total scattering cross-section of the cylinder resonator as a function of its aspect ratio $r/l$ in the region of the avoided resonance crossing between the modes TE$_{1,1,0}$ and TM$_{1,1,1}$. ({b}) Peak positions for the low- and high-frequency modes in the spectra. ({c-d})   {Evolution of the quality-factor $Q$, the peak amplitude $A$ (see Eq.~\ref{eq:peak_amplitude}), and the Fano asymmetry parameter $q$ (see Eq.~\ref{eq:Fano_asymmetry}) for the high-frequency mode.}}
      \label{fig:figure_2}
\end{figure*}

The low-frequency spectrum of the dielectric cylinder under consideration consists of {\it three types of modes}. The modes with the axial index $p=0$ and azimuthal index $n=0,1$ demonstrate a small frequency shift with changing $r/l$. They are formed mainly due to reflection from a side wall of the cylinder, and they could be associated with the Mie resonances of an infinite cylinder (see Figs.~\ref{fig:figure_1}a and \ref{fig:figure_11}a). The modes with the indices $p>0$ and $n=0,1$  demonstrate a strong shift to higher frequencies with increasing aspect ratio $r/l$. They are formed mainly due to reflection from the faces of the cylinder, and they could be associated with the Fabry-Perot modes (see Figs.~\ref{fig:figure_1}a and \ref{fig:figure_11}a). The modes with the azimuthal index $n=2,3,...$ are formed due to the wave incident on the side wall of the cylinder at the angles bigger than the total internal reflection angle, which is about 6.4 degrees for $\varepsilon_1=80$. Therefore, they are close in nature to the whispering gallery modes  (see Fig.~\ref{fig:figure_11}a) and their high $Q$ factor is explained by total internal reflection but not by destructive interference as we have for quasi-BIC. Properties of WGMs are well-studied (see, e.g., Refs.~\cite{oraevsky2002whispering,Matsko2006,ilchenko2006optical}) and further we focus on the  Mie-like (TE$_{1,1,0}$) and Fabry-Perot-like (TM$_{1,1,1}$) modes.  {Their electric field distributions are shown in Fig.~\ref{fig:figure_1}b.}        

In quantum mechanics, in the simplest case, the system with light-matter interaction is described by a sum of Hamiltonian without interaction $\hat H_0$ and an interaction potential $\hat V$ (see, e.g., Ref.~\cite{Land3}). The diagonal components of $\hat V$ are responsible for energy shift and the off-diagonal components are responsible for the coupling. The interaction results in a mixing of the light and matter states and in appearance of an {\it avoided resonance crossing}   --- the characteristic feature of the {\it strong coupling} regime~\cite{Scully1997}. 

In electromagnetism, due the fact that a resonator is an open system, description of the interaction between the modes becomes more complicated.  There are two main approaches describing the interaction between the modes in open system. The first one considers an open system  (dielectric cylindrical resonator in our case) as a closed system with non-radiating modes $|\varphi_a\rangle$ and $|\varphi_b\rangle$ interacting with a continuum of the radiation modes outside of the resonator in accord with the Friedrich-Wintgen mechanism~\cite{Friedrich1985} (Fig.~\ref{fig:figure_11}b). The difficulty of this method is to correctly define the basis of the non-radiating modes and their coupling constants with the radiation continuum. In the second approach, the resonator is primordially considered as an open non-Hermitian system, characterised by a complex eigenfrequency spectrum. In this approach, a small change of resonator shape could be described as a perturbation $\delta\varepsilon(\mathbf{r})$ playing a role of the interaction potential $\hat V$ between modes $|\varphi_a\rangle$ and $|\varphi_b\rangle$ (see Refs.~\cite{Lai1991,lai1990time}). In our case, a perturbation $\delta\varepsilon(\mathbf{r})$ is responsible for change of the aspect ratio of the cylindrical resonator (Fig.~\ref{fig:figure_11}c). This method is well-developed for quantum mechanics and electrodynamics~\cite{Ching1998,muljarov2011brillouin,Zeldovich1961,brillouin1932problemes,More1971}. It allows to find spectrum, eigenmodes, and interaction constants straightforwardly from the Maxwell's equations (see Appendix~\ref{appendix:analytics}). 

For the cylindrical resonator, the strong coupling  between the  Mie-like and Fabry-Perot-like modes is clearly manifested in the map of the SCS as avoided resonance crossing points (Figs.~\ref{fig:figure_1}c and~\ref{fig:figure_1}d). The most pronounced regions of the avoided resonance crossing are marked by red ellipses in Fig.~\ref{fig:figure_1}d. More detailed analysis shows that in the vicinity of the avoided resonance crossing, the $Q$ factor of one the coupled mode becomes very high that corresponds to the appearance of a quasi-BIC.   {The dramatical increase of the Q factor is a result of destructive interference between the modes with similar radiation patterns in far field. 
}

\subsection{Analysis of the Fano parameter}

The scattering of light by high-index dielectric particles is governed by the Mie resonances of the structure. In the case of highly-symmetric geometries, such as infinite rods~\cite{rybin2013mie,rybin2015switching}, spheres~\cite{tribelsky2016giant} or core-shell particles~\cite{kong2016fano} it was shown by means of the Mie theory that the scattering cross section represents a series of Fano resonances where each resonance can be described by the Fano formula~\cite{Fano1961,limonov2017Fano}. For other designs of resonators, e.g. finite dielectric cylinders, analytical Mie solution does not exist because the variables of the Maxwell's equations are not separable. However, description of the SCS by the Fano formula is still convenient but, in general, Fano parameters are introduced phenomenologically. The most challenging problem is determination of exact expressions for Fano parameters in concise and clear form. It is worth mentioning that the Fano formula for resonators with complicated geometries can be obtained but for special assumptions~\cite{gallinet2011ab}.

In this paper, we derive an elegant analytical solution for the finite-size cylinder scattering problem by proving rigorously that SCS of a lossless dielectric cylinder irradiated by a plane wave represents the conventional Fano formula. By this approach, we investigate the strong coupling between Mie-like and Fabry-Perot-like modes and reveal the relationship between the peculiarities of the mode spectra, particularly, emergence of quasi-BICs, and the singularities of Fano parameters.

%We study the scattering of linearly-polarized plane wave $\mathbf{E}_\mathrm{ inc}$ by the dielectric cylinder with permittivity $\varepsilon(\omega, \mathbf{r})$ without absorption. %\begin{equation}
%\sigma = \frac{4\pi c}{\omega\mathbf{E}_0^2}\mathrm{ Im}\left[\mathbf{E}_0 \cdot \mathbf{\hat E}_\mathrm{ sc}(\mathbf{n}_\mathrm{ i})\right].
%\label{eq:F1}
%\end{equation}
%Here $\mathbf{E}_0$ is the amplitude of the plane wave and $\mathbf{\hat E}_\mathrm{ sc}$ is the amplitude of scattered field calculated in the direction of incident light $\mathbf{n}_\mathrm{ i}$.
% The scattered wave is formed of resonant and non-resonant scattering amplitudes (see Supplemental Materials).

We assume harmonic time dependence of the incident field in the form $\mathbf{E}_{\text{inc}}e^{-i\omega t}$ and determine SCS through the extinction cross section which for the lossless case can be calculated using the optical theorem~\cite{evlyukhin2016optical}. The main idea of our approach is the expansion of scattered field amplitude into the sum of independent terms where each term corresponds to an eigenmode of the cylinder. This becomes possible by applying the recently developed procedure of the resonant-state expansion that allows for rigorous characterization of eigenmode spectrum of open optical resonators~\cite{doost2014resonant}. The cylinder eigenmodes (or resonant states) are treated as self-standing resonator excitations with a complex spectrum describing both the resonant frequencies  $\omega_0$ and damping rates $\gamma$. 
%Thus, we decompose the resonant contribution where each term is governed by the excitation of an individual resonant state $\mathbf{E}_\mathrm{ rs}$. 
Our straightforward but cumbersome calculations (see Supplemental Materials) show that the frequency dependence of the SCS could be rigorously described by the Fano formula and the Fano parameters could be expressed analytically through the material and geometrical parameters of the cavity:
\begin{subequations}
\begin{align}
\sigma&(\omega) = \frac{c^2}{\omega^2 |\mathbf{E}_\mathrm{inc}|^2}\left[\frac{A}{1+q^2}\frac{(q+\Omega)^2}{1+\Omega^2} + I_\mathrm{ bg}(\omega)\right],\\ 
\label{eq:peak_amplitude}
A &= c|\kappa|^2/2\gamma,\\
\label{eq:Fano_asymmetry}
q &= -\cot{\Delta},\\
 \Delta &= \arg{(\kappa)}.
\end{align}
\end{subequations}
Here $\Omega=(\omega-\omega_0)/\gamma$ is the relative frequency detuning,  $q$ is the Fano asymmetry parameter, $A$ is the smooth amplitude of the peak, $\Delta$ is the resonant phase and $I_\mathrm{ bg}$ is the background contribution describing the non-resonant scattering terms. For $\omega$ in the vicinity of one of the eigenfrequencies $\omega_0 - i\gamma$ frequency dispersion of Fano parameters can be neglected. The key parameter of the model which determines both $q$ and $A$ is the coupling coefficient between the electric field of the resonant state $\mathbf{E}_\mathrm{ rs}$ and of the incident field $\mathbf{E}_\mathrm{ inc}$
\begin{equation}
\kappa=-\frac{\omega_0^2}{c^2} \int\limits_\mathrm{ cylinder} d\mathbf{r}\ [\varepsilon_1-\varepsilon_2]\mathbf{E}_\mathrm{ rs}(\mathbf{r})\cdot \mathbf{E}_\mathrm{ inc}(\omega_0, \mathbf{r}).
\label{eq:kapp}
\end{equation}
The developed approach allows for investigation of evolution of the Fano parameters of coupled modes in parametric space (for different aspect ratio of the cylinder). As an example, we apply it to study the coupling between TE$_{1,1,0}$ and TM$_{1,1,1}$ modes which SCS spectrum in the vicinity of the avoided resonance crossing is shown in Fig.~\ref{fig:figure_2}(a). The calculated resonance positions, Q factor and Fano parameters $A$ and $q$ together with results of extraction of the same data by numerical fitting of the SCS to the Fano formula are shown in Fig~\ref{fig:figure_2}(b-e).

Analysis of Fig.~\ref{fig:figure_2} reveals strong correlation between evolution of Fano parameters and quality factor of the high-frequency band. Foremost, $q$ tends to infinity exactly at $r/l = 0.543$ where the quasi-BIC with high Q factor emerges. Next, at $r/l = 0.59$ where the SCS shows a narrow deep with a symmetric quasi-Lorentzian antiresonance and $q=0$, the peak amplitude $A$ decreases dramatically to a negligible but non-zero value. This means that mode excitation from the far field is strongly suppressed so the mode becomes almost dark. Importantly, we can claim that a quasi-BIC with high Q factor and a dark mode with vanishing intensity manifest themselves at different values of the aspect ratio. This counterintuitive result shows that quasi-BIC differs substantially from a true BIC supported by unbound structures, which is always a dark mode. While true BICs are invisible in the scattering spectrum, quasi-BICs can be tracked easily by controlling peak shapes in the SCS.

%A true BIC with infinite quality-factor supported by unbound structures also appears to be a dark mode which excitation by any far field source is forbidden. In the case of a finite-size cylinder resonator the mode which possesses the features of the quasi-BIC inherits the regime of a dark mode, but manifests itself at the different value of aspect ratio. It can be seen from the analysis of the evolution of high-frequency peak properties in Fig.~\ref{fig:figure_2}. For cylinder aspect ratio  and the mode becomes almost dark. At this point  which corresponds to the critical value of the Fano parameter $q=0$ as shown in Fig.~\ref{fig:figure_2}(e). Moreover,  Therefore, the peculiarities of $q$ are strongly related to formation of high-Q states and dark modes which are the hallmarks of a true BIC in infinite systems.

\begin{figure}
   \centering
  \includegraphics[width=0.9\linewidth]{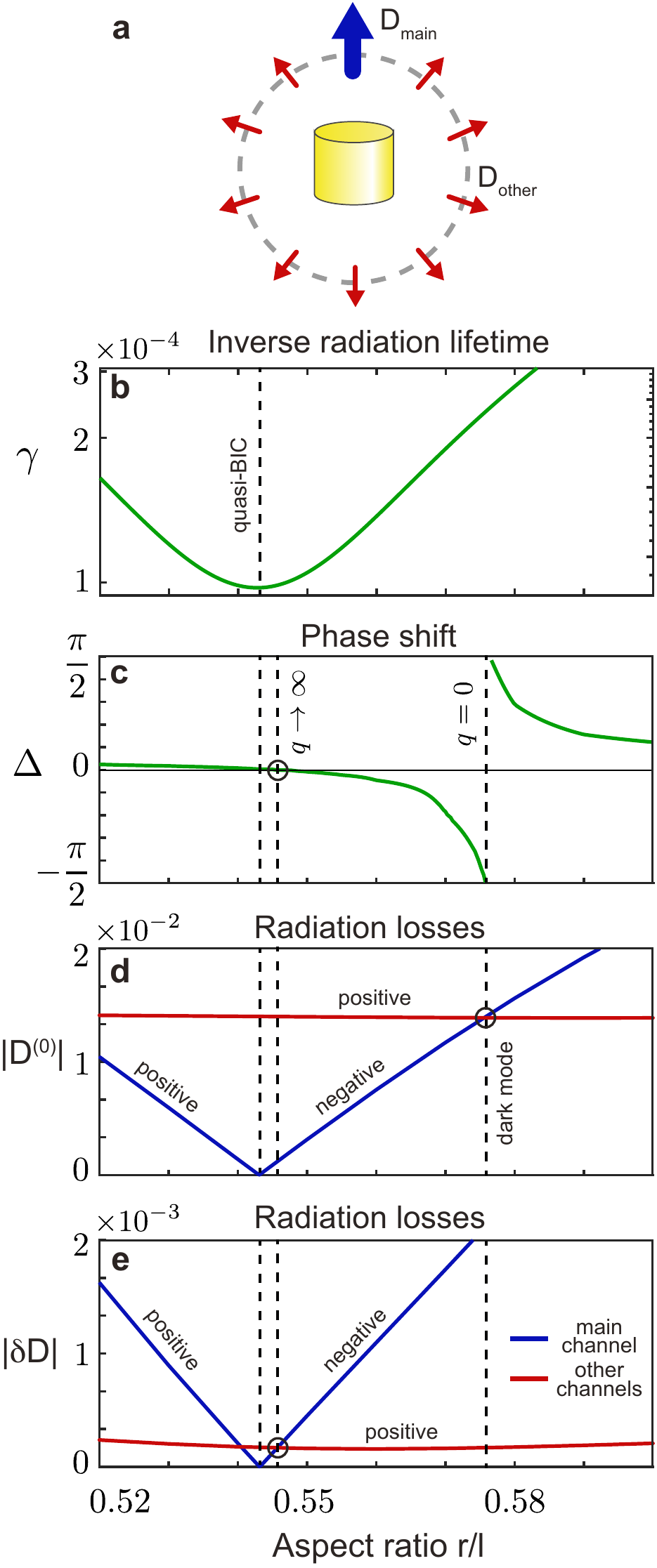} % requires the graphicx package
   \caption{ {Relationship between Fano parameter and Q-factor}. ({a}) Artistic view of open resonator which radiates into the main open channel $\mathrm{ D_{\text{main}}}$ and other minor  channels $\mathrm{ D_{\text{other}}}$. ({b}) Dependence of the inverse radiation lifetime $\gamma$ on the cylinder aspect ratio for the high-frequency mode (see Fig.~\ref{fig:figure_2}). ({c})  Dependence of the phase shift $\Delta$ on the aspect ratio $r/l$. ({d}) Dependence of the zero-order radiation amplitudes $\mathrm{ D}^{(0)}_\mathrm{ main}$ and $\mathrm{ D}^{(0)}_{\text{other}}$ on the aspect ratio $r/l$. ({e}) Dependence of the first-order corrections to the radiation amplitudes, $\delta\mathrm{ D}_\mathrm{ main}$ and $\delta\mathrm{ D}_{\text{other}}$ on the aspect ratio $r/l$.}
   \label{fig:figure_4}
\end{figure}

To gain deeper insight into a link between formation of quasi-BIC and peculiarities of Fano parameters in SCS spectra, we consider the radiation losses as a perturbation. This approach is natural and justified since we are working in the vicinity of quasi-BIC, where radiation losses are strongly suppressed.  
%Since we are focused on high-Q states radiation losses can be treated as a perturbation of modes of a closed resonator by coupling to the continuum of radiation modes. 
Remarkably, nonperturbative extension of this method is developed in Ref.~\cite{gongora2017fundamental}. In general, radiation continuum represents a set of independent channels which we label by ${\alpha}$. \textcolor{black}{For a single resonator, the independent channels can be attributed to the spherical multipoles. The details of the multipole analysis of quasi-BICs are provided in Sec.~\ref{sec_mult_an}.} %For most of the cylinder modes their radiation pattern is governed by a single dominant channel, which we denote as the main one while the rest small amount of power goes via other channels (see Fig.~\ref{fig:figure_4}a). 
The coupling amplitude $\mathrm{ D}_{\alpha}$ between the resonant state and the radiation continuum mode $\mathbf{E}_{\alpha}$ is given by (see Supplementary materials)
\begin{equation}
\mathrm{ D}_{\alpha} = -\frac{\omega_0^2}{c^2} \int\limits_\mathrm{ cylinder} d\mathbf{r}\ [\varepsilon_1-\varepsilon_2]\mathbf{E}_\mathrm{ rs}(\mathbf{r})\cdot \mathbf{E}_{\alpha}(\omega_0, \mathbf{r}).
\end{equation}

According to the the reciprocity theorem, the same amplitudes  $\mathrm{ D}_\alpha$ determine coupling with the incident field~\cite{suh2004temporal}.
Following the perturbation theory, we represent each resonant state $\mathbf{E}_\mathrm{ rs}$ as sum of a closed resonator mode $\mathbf{E}^{(0)}$ and first order correction $i\delta\mathbf{E}$ responsible for the radiation. Therefore, the coupling constant can be also expanded as $\mathrm{ D}_\alpha = \mathrm{ D}^{(0)}_\alpha + i\delta\mathrm{ D}_\alpha$. The inverse radiation lifetime $\gamma$ of the resonant state can be calculated as the sum of radiation losses into all radiation channels (like the Fermi's golden rule in quantum mechanics): 
%The radiation intensity is governed by the same coupling amplitude $D$ because of the reciprocity theorem~\cite{suh2004temporal}, thus 
\begin{equation}
2\gamma = c\sum_{\alpha} |\mathrm{ D}_{\alpha}|^2.
\label{eq:Fermi}
\end{equation}
Furthermore, the coupling coefficient $\kappa$ between the resonant state and the incident field (see Eq.~\ref{eq:kapp}) can be decomposed into a series of independent contributions of all channels (see Supplemental materials). For most of the eigenmodes their radiation is mainly determined by a single dominant channel, which we denote as $\mathrm{ D}_{\text{main}}$. The rest small amount of power goes via other channels, which we denote as $\mathrm{ D}_{\text{other}}=\sum_{\alpha}'\mathrm{ D}_\alpha$ (see Fig.~\ref{fig:figure_4}a). The sum is taken over all radiation channels except the main one. In these notations, the critical behaviour of the Fano asymmetry parameter $q$ is determined by the following simple conditions:
%Remarkably, only the ratio of $\mathrm{ D}^{(0)}$ and $\delta\mathrm{ D}$ determines the value of $\kappa$ and, as a consequence, $q$ and the shape of peaks observed in the scattering spectra. 
%For most of the cylinder modes their radiation pattern is governed by a single dominant channel, which we denote as the main one while the rest small amount of power goes via other channels (see Fig.~\ref{fig:figure_4}a). 
%More rigorously, critical behaviour of Fano asymmetry parameter $q$ is determined by the conditions
\begin{subequations}
\begin{align}
\label{eq:law_a}
\delta\mathrm{ D}_\mathrm{ main} &+ \delta\mathrm{ D}_{\text{other}} = 0, \qquad q\rightarrow \infty, \\
\mathrm{ D}^{(0)}_\mathrm{ main} &+ \mathrm{ D}^{(0)}_{\text{other}} = 0,  \, \ \qquad q = 0.
\label{eq:law_b}
\end{align}
\label{eq:D} 
\end{subequations}

\begin{figure}
   \centering
  \includegraphics[width=0.9\linewidth]{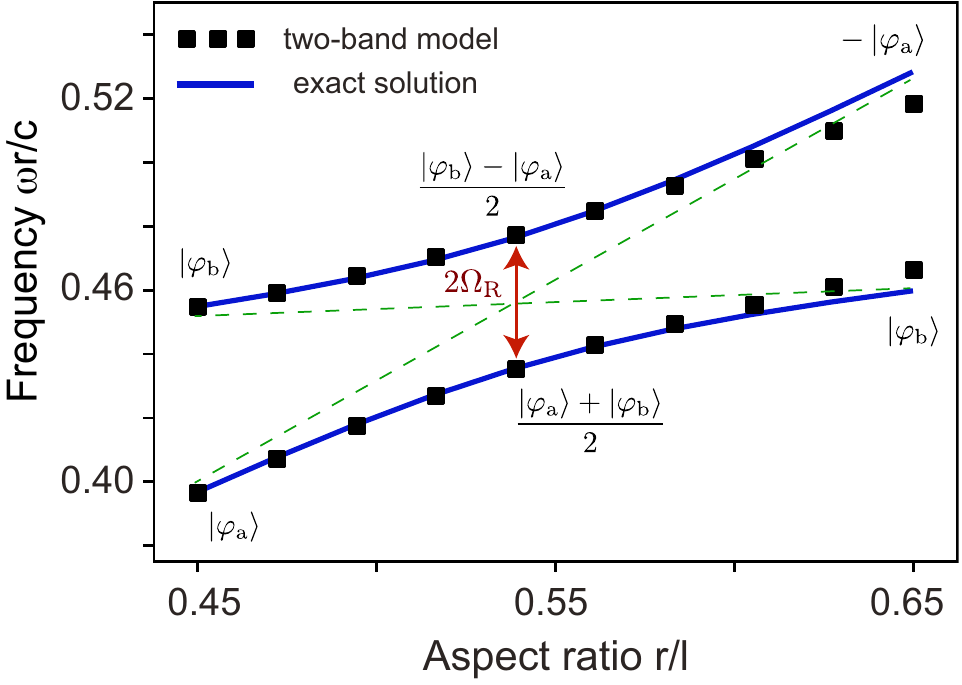} % requires the graphicx package
   \caption{ {Two-band approximation of strong mode coupling.} Comparison of the exact solution and approximate two-band model of strong coupling between the modes TE$_{1,1,0}$ and TM$_{1,1,1}$. Green dashed lines are guide for eyes. }
   \label{fig:figure_5}
\end{figure}
Inspection of Eq.~\ref{eq:Fermi} and Eqs.~\ref{eq:D} reveals the crucial role of the main channel in both formation of high-Q modes and critical behavior of $q$. If radiation to the main channel is completely suppressed ($\mathrm{ D}^{(0)}_\mathrm{ main}+\delta\mathrm{ D}_\mathrm{ main}=0$) we get a quasi-BIC, characterized by minimal radiation losses. However, the Fano asymmetry parameter $q$ tends to infinity not exactly at quasi-BIC, but very close to it, when condition \ref{eq:law_a} is fulfilled. The dark mode ($q=0$) is formed under condition \ref{eq:law_b}, when the radiation into the main and the rest channels compensate each other in the first approximation. 

%characterized by symmetric antiresonance shape of SCS spectru-Lorentz shape.... quasi-BIC. Since radiation to other channels is weak conditions of Eqs.~\ref{eq:D} should be satisfied almost simultaneously with emergence of quasi-BICs. 
For deeper understanding we focus on the particular example and investigate evolution of coupling coefficients $\mathrm{ D}_\alpha$ for the high-frequency band of the avoided resonance crossing between TE$_{1,1,0}$ and TM$_{1,1,1}$ modes. In this case the main channel represents electric dipole radiation and other channels are dominated by the magnetic quadrupole radiation. The  comparison of the evolution of the inverse radiation lifetime $\gamma$, phase $\Delta$ and the amplitudes $\mathrm{ D}^{(0)}$ and $\delta\mathrm{ D}$ for different channels with respect to $r/l$ is shown in Figs.~\ref{fig:figure_4}b-e, respectively.

Figures~\ref{fig:figure_4}d and~\ref{fig:figure_4}e show that the radiation to the main channel is suppressed in virtue of coupling between TE$_{1,1,0}$ and TM$_{1,1,1}$ modes and it is completely vanished at $r/l = 0.543$ where the quasi-BIC emerges.
Importantly, in this regime $\delta\mathrm{ D}_{\text{main}}$ nulls simultaneously with $\mathrm{ D}^{(0)}_\mathrm{ main}$ since both of them are proportional to the rate radiation into the main channel. Under further change of the aspect ratio $r/l$ the amplitudes $\mathrm{ D}_{\text{main}}^{(0)}$ and $\delta\mathrm{ D}_{\text{main}}$  become negative and continue to decrease. At $r/l = 0.546$, the perturbation $\delta\mathrm{ D}_{\text{main}}$ compensate $\delta\mathrm{ D}_{\text{other}}$ and $q$ goes to infinity (Eq.~\ref{eq:law_a}).  For further increase the aspect ratio, the unperturbed amplitude $\mathrm{ D}^{(0)}_{\text{other}}$ of the non-dominant channels becomes exactly opposite to $\mathrm{ D}^{(0)}_\mathrm{ main}$ at $r/l = 0.575$. Thus, conditions \ref{eq:D}b are satisfied and $q=0$. Therefore,  the quasi-BIC and dark mode are not appeared at the same value of the aspect ratio because of contribution of other non-dominant channels of radiation losses.

\subsection{Multipole analysis \label{sec_mult_an}}

In order to gain deeper insight into the physics of quasi-BIC in a single resonator we illustrate cancelation of its radiation losses through the dominant channel in term of multipoles. The far-field of a single resonator could be expanded into a multipole series of vector spherical harmonics.
%\footnote{For metasurfaces, more natural basis for far-field expansion is plane waves, which correspond to open diffraction channels of periodic structure. The significant difference between the single resonator and periodic structures is the number of radiation channels. It is important that for periodic structures for any frequency we have finite number of open diffraction channels but for single non-spherical resonator this number is always infinite.}.
Each harmonic plays a role of an independent radiation channel. One-to-one  correspondence between eigenmodes and spherical multipoles can be established only for spherical resonators. Any mode of other resonators is always contributed by the infinite number of multipoles. However, in this infinite series it is possible to distinguish the dominant term (dominant channel) making the main contribution to the radiated power. 

\begin{figure}
   \centering
  \includegraphics[width=0.9\linewidth]{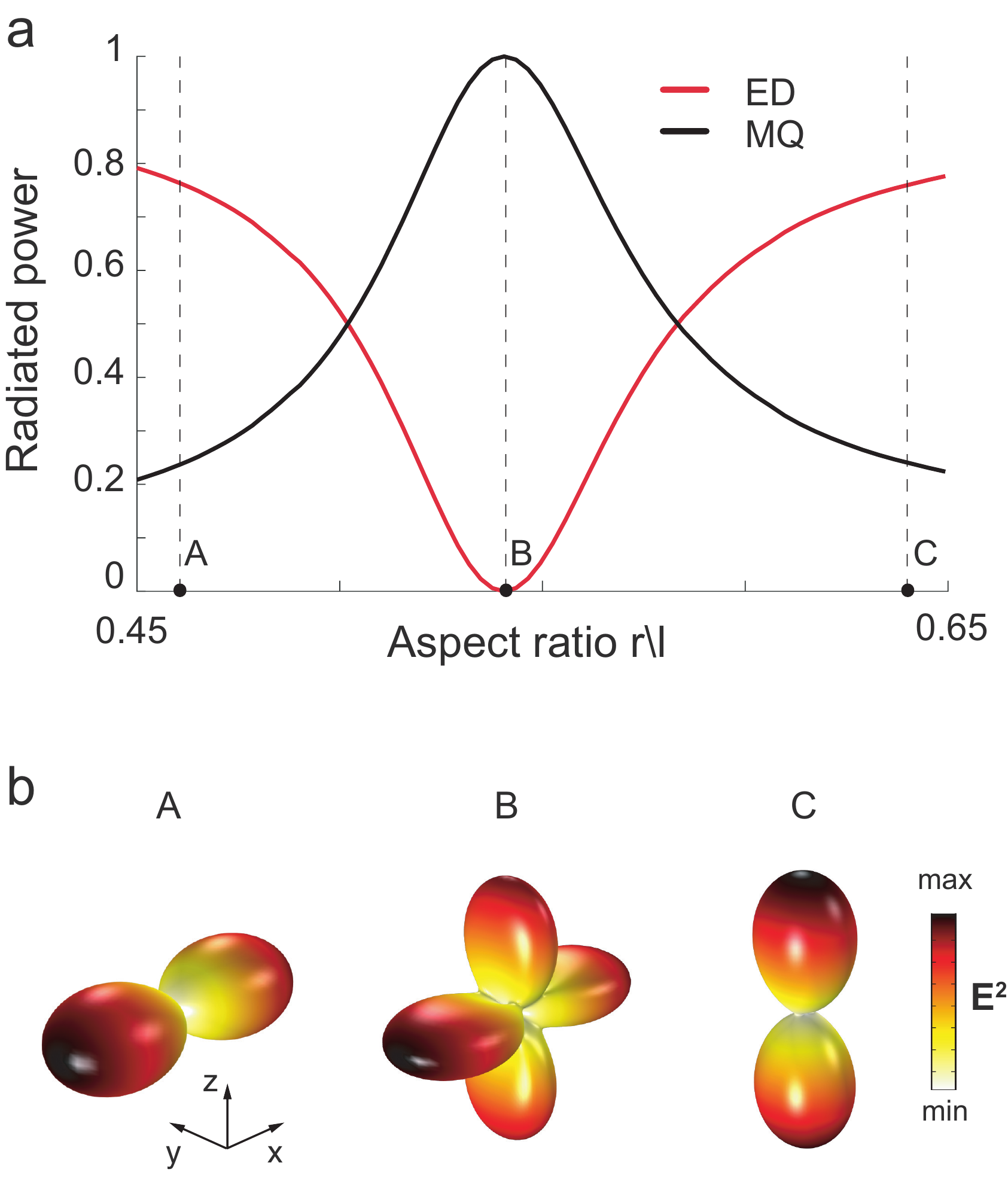} % requires the graphicx package
   \caption{ Multipole decomposition for TM$_{1,1,1}$ mode. (a) Contribution of the electric dipole and magnetic quadrupole to the radiated power of TM$_{1,1,1}$ mode.  (b) Far-field radiation patterns of TM$_{1,1,1}$ mode for different aspect ratios. Panel B corresponds to the quasi-BIC.}
   \label{fig:figure_multiples}
\end{figure}

It is possible to show using the group symmetry analysis that for the mode TM$_{1,1,1}$ the main contribution to the radiation energy is given by electric dipole moment and the rest part of energy is mainly radiated through the magnetic quadrupole moment. For the aspect ratio $r/l$ of the quasi-BIC, radiation through the dipole channel becomes negligible and the dominant radiation channel is the magnetic quadrupole [see Fig.~\ref{fig:figure_multiples}(a)] and the radiation pattern changes dramatically [see Fig.~\ref{fig:figure_multiples}(b)].  It is possible to show that all other quasi-BICs demonstrate similar bahaviour in far-field~\cite{Multipole_analysis}.  Therefore, quasi-BICs in single resonators are manifested not only in the scattering spectra as a singularity of the Fano asymmetry parameter but in the far-field since the radiation pattern changes dramatically.

\subsection{Two-band model of strong mode coupling}

In this section we discuss the mechanism of strong coupling between modes in a cylindrical resonator and the reason of complete suppression of radiation to the main channel. As it was shown in the previous section, mode coupling is realised not in real, but in parametric space. It  means that the cylinder aspect ratio represents the parameter determining the strength of interaction between modes. Therefore, coupling between the modes is governed by the perturbation of cylinder permittivity (see Fig.~\ref{fig:figure_11}c). 

Generally, this perturbation mixes all resonant states of the cylinder. New resonant states of the perturbed resonator can be found by means of the resonant-state expansion (see Appendix~\ref{appendix:analytics}) which is a special  rigorous technique that allows for careful investigation of spectrum of open systems. However, in the vicinity of an avoided resonance crossing the interaction between modes involves only two of them. Therefore, the general form of the resonant-state expansion can be reduced to a two-band model.

\begin{figure}[t]
   \centering
   \includegraphics[width=0.9\columnwidth]{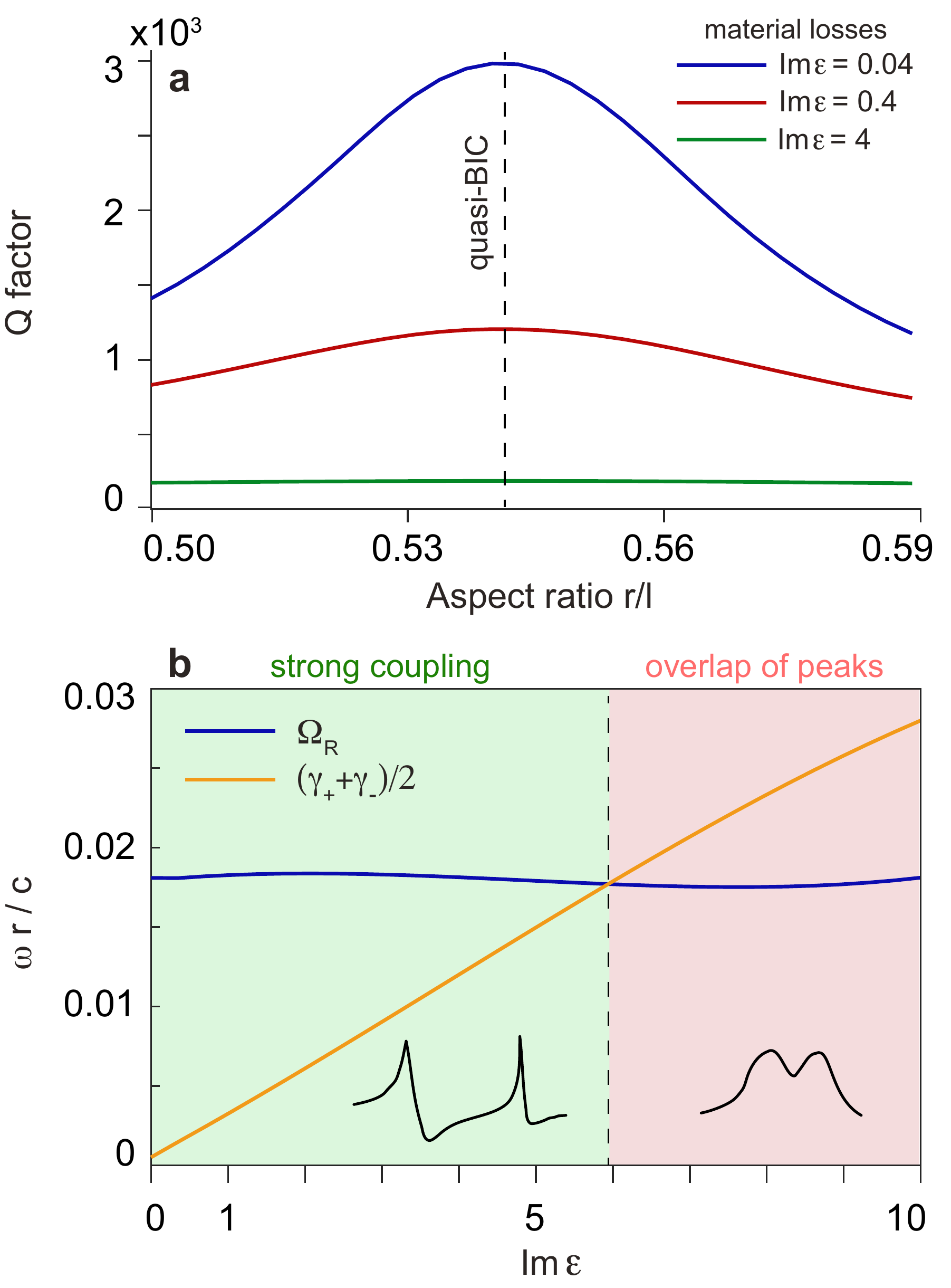} % requires the graphicx package
   \caption{{Effect of material losses on regime of strong coupling and quasi-BIC}  ({a}) Dependence of the total quality-factor $Q$ on the aspect ratio for various levels of material losses. ({b}) Dependence of Rabi frequency and sum of half linewidths of the coupled modes on the level of material losses. Insets visualise ratio between $\Omega_R$ and linewidths. Here, $\gamma_+$ and $\gamma_-$ are the damping rates of the modes of the diagonalized Hamiltonian~\eqref{eq:FWH}.}
   \label{fig:figure_losses}
\end{figure}

\begin{figure*}[t]
   \centering
   \includegraphics[width=0.99\linewidth]{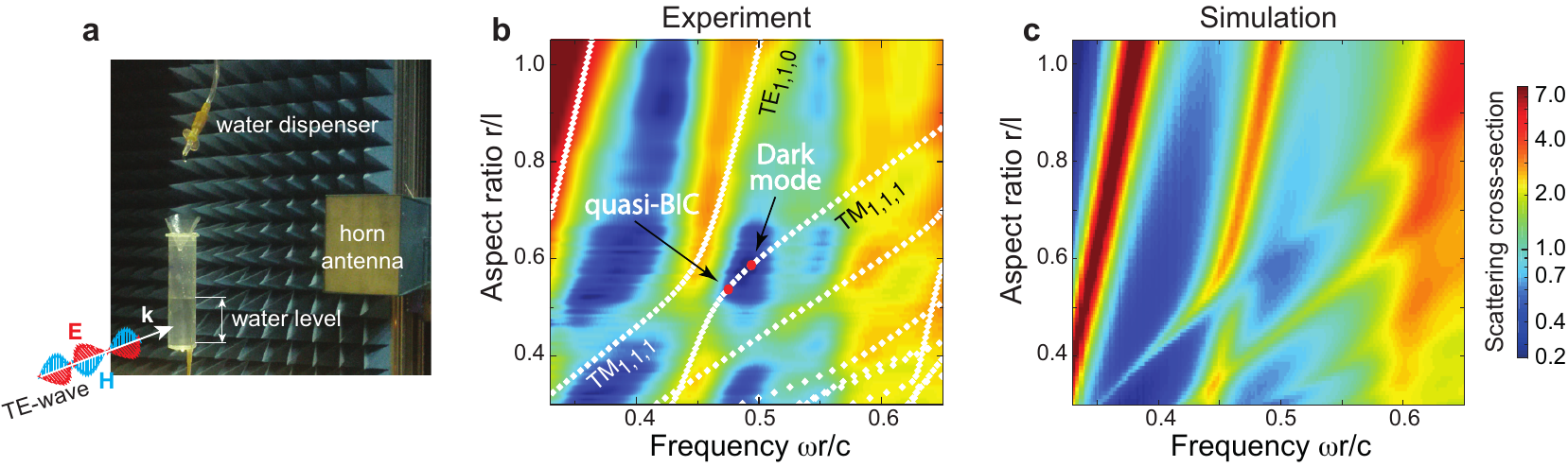} % requires the graphicx package
   \caption{{Experimental results.}  ({a}) Experimental setup for the measurement of SCS spectra of the cylindrical resonator filled with water depending on its aspect ratio $r/l$ and frequency $\omega r /c$. ({b}) Measured SCS map demonstrating the avoided crossing regime between TE$_{1,1,0}$ and TM$_{1,1,1}$ resonances. The circles are the real part of eigenfrequencies obtained from the resonant state expansion method for a dielectric cylinder with the permittivity $\varepsilon_1=80$  embedded in air ($\varepsilon_2=1$). ({c}) Calculated SCS map of the cylindrical resonator filled with water depending on the frequency $\omega r /c$ and aspect ratio $r/l$. The frequency dispersion of the water permittivity is taken from Ref.~\cite{kaatze1989complex}.}
   \label{fig:figure_7}
\end{figure*}

We start from the two of uncoupled cylinder eigenfunctions $\ket{\varphi_\mathrm{ a}}$ and $\ket{\varphi_\mathrm{ b}}$ with eigenfrequencies $\omega_\mathrm{ a}-i\gamma_\mathrm{ a}$ and $\omega_\mathrm{ b}-i\gamma_\mathrm{ b}$, respectively. They undergo strong coupling and the perturbed eigenfunction with eigenfrequency $\overline\omega$ represents their linear combination $\ket{\varphi}=C_\mathrm{ a}\ket{\varphi_\mathrm{ a}}+C_\mathrm{ b}\ket{\varphi_\mathrm{ b}}$, where coefficients $C_\mathrm{ a,b}$ are, in general, complex. The two-band model can be written as a generalized eigenvalue problem~\cite{doost2014resonant}

\begin{equation}
\begin{bmatrix}
\omega_\mathrm{ a}\!-\!i\gamma_\mathrm{ a} & 0\\
0 & \omega_\mathrm{ b}\!-\!i\gamma_\mathrm{ b}
\end{bmatrix}\begin{bmatrix}
C_\mathrm{ a}\\
C_\mathrm{ b}
\end{bmatrix}=\overline\omega
\begin{bmatrix}
1\!+\!V_\mathrm{ aa} & V_\mathrm{ ab}\\
V_\mathrm{ ba} & 1\!+\!V_\mathrm{ bb}
\end{bmatrix}
\begin{bmatrix}
C_\mathrm{ a}\\
C_\mathrm{ b}
\end{bmatrix}.
\label{eq:FWH}
\end{equation}
Here the perturbation is determined by the symmetric matrix with complex-valued elements  which makes the problem non-Hermitian
\begin{equation}
V_\mathrm{ ij} = \frac{1}{2}\int d\mathbf{r}\ \delta\varepsilon(\mathbf{r})\mathbf{E}_\mathrm{ rs}^\mathrm{ i}(\mathbf{r})\cdot\mathbf{E}_\mathrm{ rs}^\mathrm{ j}(\mathbf{r}), \ \ \ \ \mathrm{ i,j}=\mathrm{ a,b}.  
\label{eq:V}
\end{equation}
Figure~\ref{fig:figure_5} demonstrates perfect coincidence of the exact dispersion of two coupled modes and the two-band model results. Since both modes possess sufficiently low radiative losses, when we change cylinder aspect ratio they couple predominantly in the near-field region inside the resonator. Therefore, when the coupling is maximal, the coefficients become  $C_\mathrm{ a} = 1$, $C_\mathrm{ b} = \pm 1$ for low- and high-frequency modes in Fig.~\ref{fig:figure_5}, respectively. We recall, that $\ket{\varphi_a}$ and $\ket{\varphi_b}$  are characterized by similar far-field pattern as they have the same mode symmetry with respect to azimuthal direction and inversion symmetry of structure (see Methods). This means that far-field distribution of radiation of the high-frequency mode in the strong coupling regime is almost suppressed, i.e. $\left[\ket{\varphi_b}-\ket{\varphi_a}\right]|_{r\rightarrow \infty}\simeq 0$. Therefore, the radiation to the main channel for this value of $r/l$ becomes completely forbidden which explains to formation of high-Q quasi-BIC mode and the dark mode discussed in the previous section.

Conventional method to characterize strength of mode coupling is the Rabi frequency $\Omega_\mathrm{ R}$, which is a half of minimal distance between the dispersion curves of coupled modes (see Fig.~\ref{fig:figure_5}). For the two resonances to be spectrally separable, the minimum mode-splitting needs to be greater than the sum of the half linewidths of the modes, which is a necessary condition to observe strong coupling~\cite{zhang2017room}. For the avoided resonance crossing under consideration $\Omega_\mathrm{ R}r/c=0.018$ which is 35 times higher than sum of the half linewidths which clearly manifests the strong coupling of modes.

\subsection{Effect of material losses}

For the cylindrical resonator analyzed above we neglect the material losses and take into account only the radiative ones. Here we analyzed the effect of material losses on  { the quality factor of quasi-BIC and on the conditions of strong coupling}. Figure \ref{fig:figure_losses}a shows the dependence of the total quality factor $Q$ on the aspect ratio for the high-frequency mode in the vicinity of the avoided crossing regime between the modes TE$_{1,1,0}$ and TM$_{1,1,1}$ at different material loss level. One can see that the $Q$ factor strongly depends on the material losses and could be substantially decreased. In the presence of material losses the total $Q_\text{tot}$ factor could be estimated  as
\begin{equation}
Q^{-1}_\text{tot}=Q^{-1}_\text{rad}+Q^{-1}_\text{mat}.
\end{equation}
Here $Q_\text{rad}$ and $Q_\text{mat}$ are responsible for the radiative and material losses, respectively. Therefore, the results obtained for the lossless cavity are justified and $Q_\text{tot}\approx Q_\text{rad}$ if the radiative losses are dominant ($Q_\text{rad}>Q_\text{mat}$). 

 {
Since material losses decrease the Q factor, they affect the strength of mode coupling as well. Figure \ref{fig:figure_losses}b shows how the Rabi frequency $\Omega_\text{R}$ and the sum of the half linewidths change depending on the level of material losses. Strong coupling regime breaks when the resonance become spectrally inseparable which is realized for $\mathrm{ Im}\ \varepsilon = 6$. Therefore, even for relative high absorption the strong coupling can be realized, which is extremely useful for the experimental measurements described in the next Section.
}

\subsection{Experimental results}

Finally, we perform the experimental study to demonstrate the existence of the avoided crossing regime between the TE$_{1,1,0}$ and TM$_{1,1,1}$ resonances in the microwave frequency range. In the experiment, the plastic cylindrical vessel filled with water is placed in the middle between two antennas. The aspect ratio of the cylindrical resonator is defined by the amount of water. The photo of the experimental setup is shown in Fig.~\ref{fig:figure_7}a (see Appendix~\ref{appendix:experiment} for details). The resonator is excited by TE polarized electromagnetic wave incident perpendicular to the cylinder axis $z$ (see Fig.~\ref{fig:figure_7}a).  The measured dependence of the SCS of the cylindrical resonator depending on its aspect ratio is shown in Fig.~\ref{fig:figure_7}b. The results of the numerical simulations taking into account the losses in water are shown in Fig.~\ref{fig:figure_7}c. One can see that the experimental positions of the resonances are in a good agreement with the real part of eigenfrequencies (marked by white circles) calculated using the resonant state expansion method (see Methods for details).  In spite of losses in water, which broaden the resonances, the avoided crossing regime between the TE$_{1,1,0}$ and TM$_{1,1,1}$ modes and suppression of SCS clearly manifest themselves for the aspect ratio in the range of $0.5<r/l<0.6$. Discrepancies between the measured and calculated maps of SCS could be explained by not perfect plane wave radiated by a horn antenna and parasitic scattering from the auxiliary equipment (holder of the resonator and plastic cylindrical vessel).

\section{Discussion}

 {

As we mentioned above, a true BIC is mathematical abstraction  and it is not practically implementable. \textcolor{black}{However for periodic photonic structures with quite large number of periods, the radiation of high-Q states could be almost suppressed being much less that other loss mechanisms in the system. Such states are closest to the true BICs.}
%However, we still can achieve high-Q states by suppressing the radiation losses through destructive interference and making them negligible with respect to other loss mechanisms. Such modes with high quality factor can be treated as  {\em quasi-BICs}. Such practical BICs realized in periodic photonic structures and arrays of dielectric resonators are studied recently by other groups.
Here we demonstrate that the radiation losses can be substantially suppressed,  being much smaller than other losses, via BIC-inspired mechanism even in a single isolated resonator. Therefore, BIC in a finite size periodic structure and quasi-BIC in a single resonator could be indistinguishable in practice if their radiation losses will be strongly suppressed. We \textcolor{black}{believe} that the proposed concept of quasi-BIC in a single resonator is more favourable for \textcolor{black}{compact} nanophotonic applications and much easier implementable than other designs.

%Since a quasi-BIC is characterized by a very sharp resonance, conventional spectroscopy methods demand highly-sensitive equipment to observe it. The developed theory shows that a quasi-BIC could be recognized by naked eye since the asymmetry of the Fano resonance is easly distinguishable without fitting. 

The developed theory predicts that the shape and amplitude of the SCS spectra represents an unambiguous indicator of the quasi-BIC. Namely, both regimes of $q = 0$ and $q\rightarrow \infty$ describe important features inherent to a true BIC. The latter condition $q\rightarrow \infty$ practically coincides with the emergence of a quasi-BIC. Therefore, a quasi-BIC could be recognized in the experimental spectra by the naked eye since the asymmetry of the Fano resonance is easily distinguishable without fitting. 

The difference in the aspect ratios corresponding to quasi-BIC and singularity of $q$ is determined by the magnitude of radiation losses to non-dominant channels. The difference becomes negligible with increase of the permittivity of the resonator. Even for high-index dielectric nanostructures in the visible and near-IR range, i.e. $\varepsilon~\sim10-12$, the relative difference between aspect ratios corresponding to the singular $q$ and the maximal Q factor is less than $5\%$. The predicted strong coupling regime and peculiarities of the Fano parameters could be observed in a wide spectral range from the visible to centimetre wavelengths. These results lift the veil on the nature of quasi-BICs in single resonators and emphasize profound relationship between the shape of Fano peak and emergence of bound states in the continuum. 

\textcolor{black}{Remarkably, that for BICs in periodic photonic structures, the non-dominant channels  are absent  because the radiation continuum is discretized~\cite{bulgakov2017propagating}.} Thus, the leaky modes interact with only one channel which in the simplest case represents the zero-order diffraction~\cite{zhen2014topological}. Therefore, leaky states can transform to true BICs, when the radiation to the main channel is completely forbidden. Since for true BIC, the system of Eqs.~\ref{eq:D} is satisfied simultaneously, the Fano parameter $q$ becomes ill-defined, which corresponds to the collapse of Fano resonance. Furthermore, true BIC is always a dark mode, which can be easily understood from energy consumption arguments --  BIC does not radiate at all, thus we are not able to pump it. But for finite-size dielectric cylinders the collapse of Fano resonance is not manifested, instead of this Fano parameter monotonically evolves as cylinder aspect ratio changes and consistently passes the values $q=\pm \infty$ and $q=0$ where mode exhibits features of a true BIC state.

 {Recently, the study of resonant dielectric nanostructures has been established as a new research direction in modern nanoscale optics and metamaterial-inspired nanophotonics, due to their optically-induced electric and magnetic Mie resonances~\cite{kuznetsov2016optically, baranov2017all}. However, the Q factor of Mie resonances is about tens that is far from the values achieved in WGM resonators, photonic crystal or Bragg cavities. The proposed mechanism of strong mode coupling in single high-index dielectric resonators accompanied by emergence of quasi-BIC helps to substantially extend functionality of all-dielectric nanophotonics. This opens new horizons for active and passive nanoscale metadevices including low-threshold nanolasers, biosensors, on-chip parametric amplifiers, and nanophotonics quantum circuits.}

%In centimetre wavelength range, the material $Q$ factor about 5$\times10^4$ is achievable for a variety of ceramics~\cite{ohasto2003millimeter, reaney2006microwave}. For crystalline silicon, the material $Q$ factor taking account scattering on roughness could reach $10^{6}$ (see, e.g., Ref.~\cite{Miller2017}). Therefore, the regime of quasi-BIC could be realized in a broad spectral range from visible to microwave. 

}

\section{Conclusion}
We have demonstrated that a subwavelength homogeneous dielectric resonator can support strongly interacting modes. We have shown that the strong coupling regime is accompanied by the formation of a quasi-BIC when the radiative losses are almost suppressed due to the Friedrich--Wintgen destructive interference. The analysis of the scattering cross-section reveals an abrupt change of the Fano asymmetry parameter from minus to plus infinity in the vicinity of quasi-BICs. Therefore, quasi-BIC manifests itself in scattering spectra by the symmetric Lorentzian shape. \textcolor{black}{Appearance of quasi-BIC is accompanied by drastic change of far-field radiation pattern explained by suppression of the radiation through the dominant multipole.}  This singularity could be used as an indication of quasi-BIC. In contrast to true BIC, the Fano-resonance feature for quasi-BIC does not vanish completely since it is not completely decoupled from the radiation continuum. We have confirmed our theoretical results in microwave experiment by using a cylindrical resonator filled with water. Our results open new horizons for active and passive optical nanodevices including efficient biosensors, low threshold nanolasers, perfect filters, waveguides, and nanoantennas.

\acknowledgements 

We acknowledge fruitful discussions with H. Atwater, I.V.~Shadrivov, P.A.~Belov, A.N.~Poddubny, A.~Polman, and A.~Moroz. The numerical calculations were performed with a support of the Ministry of Education and Science of the Russian Federation (Project 3.1500.2017/4.6) and  the Australian Research Council. The experimental study of the cylinder scattering cross-section in microwave frequency range has been supported by the Russian Science Foundation (17-79-20379). The analytical calculations with resonant state expansion method have been performed with a support of the Russian Science Foundation (17-12-01581). A. A. B., K. L. K. and Z. F. S. acknowledge the support from the Foundation for the Advancement of Theoretical Physics and Mathematics "BASIS" (Russia).

\appendix
\section{Experimental approach \label{appendix:experiment}}

The sample used for experimental study of SCS is a hollow plastic cylinder opened from the top with the radius of $r=20.25$~mm and high of $l=160$~mm. The thickness of the cylinder wall is $w=1.5$~mm. As a dielectric material to fill the cylinder we employed a distilled water that is characterized by permittivity of  $\varepsilon_1 \approx 80$ at room temperature~\cite{kaatze1989complex}. Forward scattering measurements have been performed in an anechoic chamber~\cite{Larsson}. A pair of wideband horn antennas (TRIM 0.75 GHz to 18 GHz; DR) have been positioned facing each other at a distance of 4 m with the sample placed  at the midpoint, see Fig.~\ref{fig:figure_7}a. The measurement uses a two ports vector network analyzer (VNA) Agilent E8362C transmitting a continuous wave. The first antenna has been connected to the first port of the VNA and provided a near plane-wave excitation in the frequency range of 0.8-5 GHz. The second horn antenna connected to the second port of the VNA has been employed as a receiver. The frequency range of  0.8-5 GHz has been swept using 10001 frequency points. Eight such sweeps are averaged for each of the sample measurement, background measurement, and calibration measurement. Calibration measurements have been performed using a metal sphere with the radius of 7.5 mm. During the sample measurements we added the water to the cylinder changing its aspect ratio $r/l$ from 0.125 to 2.5 with the average step of 0.01. The optical theorem was used to calculate the scattering cross-section from the imaginary part of the measured forward scattering amplitude~\cite{bohren2008absorption}.  {To suppress the effects of multiple reflections between the sample and the antennas, the post-processing of measured data by means of time-domain gating was employed~\cite{larsson2008extinction}.}

\section{Analytical model \label{appendix:analytics}}

We calculate the spectrum of complex eigenfrequencies of a dielectric cylindrical resonator by applying the rigorous perturbative method, the resonant-state expansion~\cite{muljarov2011brillouin}. We expand the fields $\mathbf{E}_{j}$ $(j\!=\!n,k,p)$ of eigenmodes of the cylindrical resonator over the eigenfunctions $\mathbf{E}_\alpha^{(0)}$ of a homogeneous dielectric sphere with the same value of permittivity as for the cylindrical resonator 
\begin{equation}
\mathbf{E}_j=\sum\limits_{\alpha}{b_\alpha^j}\mathbf{E}_\alpha^{(0)},
\end{equation}
where $\mathbf{E}_\alpha^{(0)}$ satisfies the Maxwell's equations with boundary conditions of outgoing waves

\begin{equation}
\nabla\!\times\!\nabla\!\times\!\mathbf{E}_\alpha^{(0)}=\varepsilon(\mathbf{r})\frac{\omega_\alpha^2}{c^2}\mathbf{E}_\alpha^{(0)}.
\end{equation}

Resonants states $\mathbf{E}_j$ satisfy the perturbed equation

\begin{equation}
\nabla\!\times\!\nabla\!\times\!\mathbf{E}_j=\left[\varepsilon(\mathbf{r})+\delta\varepsilon(\mathbf{r})\right]\frac{\Omega_j^2}{c^2}\mathbf{E}_j,
\end{equation}
where $\delta\varepsilon(\mathbf{r})$ is a perturbation that transforms a sphere into an inscribed cylinder.

The problem is non-Hermitian because of outgoing boundary conditions. Therefore, the eigenvectors grow exponentially at large distances, their proper normalization deviates from the standard Hermitian anzatz~\cite{muljarov2011brillouin}. However, $\mathbf{E}_\alpha^{(0)}$ form a complete set inside the region of a dielectric sphere and we use them as a basis.

\begin{figure}
   \centering
   \includegraphics[width=0.95\linewidth]{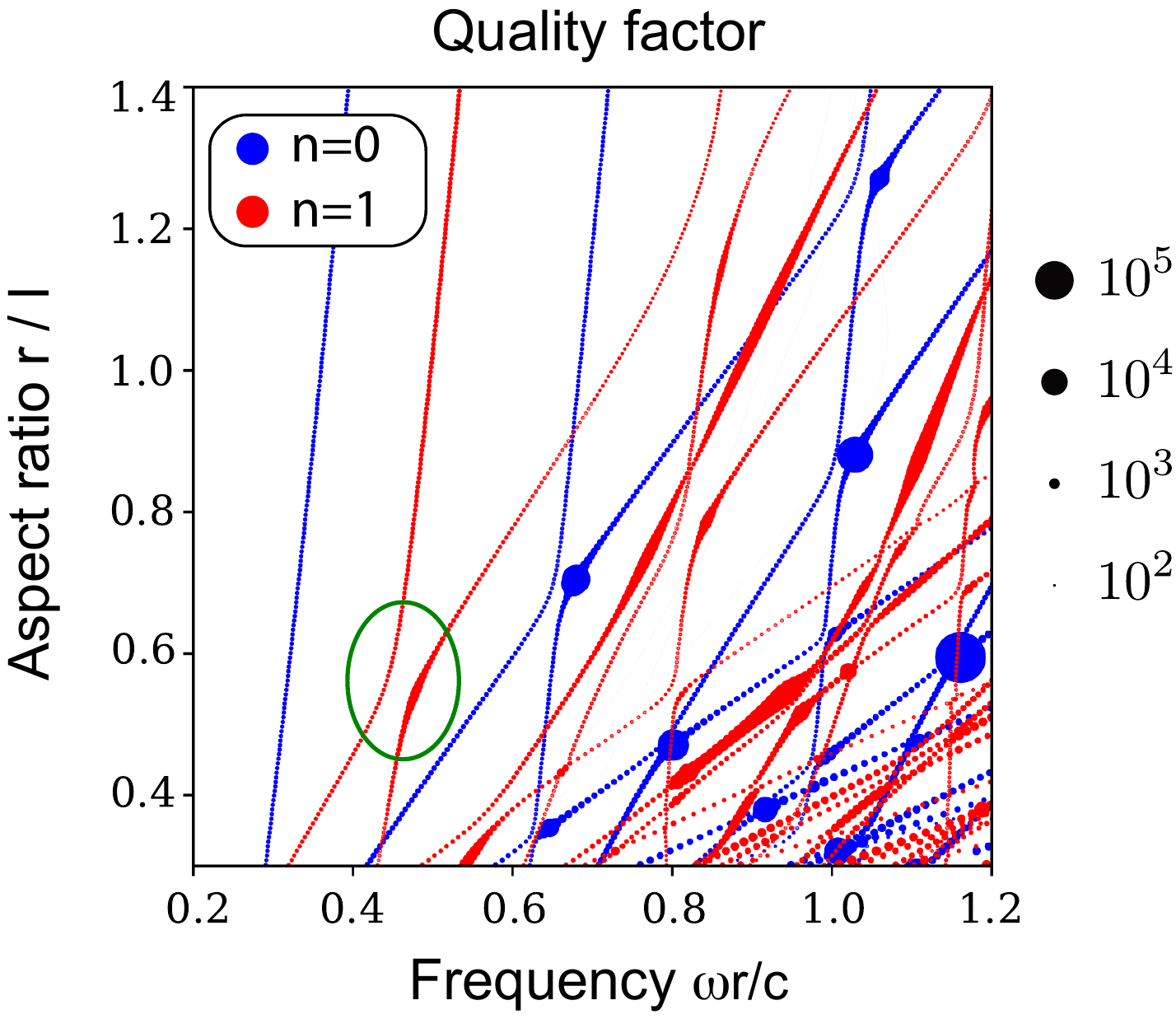} % requires the graphicx package
   \caption{{Complex spectrum of eigenmodes.} The spectrum is shown for the  modes with the azimuthal index $n=0$ (red dotted lines) and $n=\pm1$ (blue dotted lines), which are even with respect to ($z\rightarrow -z$) symmetry. Dot sizes are proportional to $Q$ factor.  The region of the avoided crossing between modes TE$_{1,1,0}$ and TM$_{1,1,1}$ is marked by the green ellipse. Calculations are performed by using the resonant-state expansion method.}
   \label{fig:figure_6}
\end{figure}

The problem is reduced to the matrix equation~\cite{muljarov2011brillouin}
\begin{equation}
\frac{1}{\omega_\alpha}\sum_\beta\left[\delta_{\alpha\beta}+V_{\alpha\beta}\right]b^j_\beta = \frac{1}{\Omega_j}b^j_\alpha,
\end{equation}
with the elements of the perturbation matrix $V_{\alpha\beta}$ were defined in Eq.~\ref{eq:V}. We should note, that here the operator $\hat V$ is responsible for the transformation of the sphere into the cylinder. Thus, this operator $\hat V$  differs from those defined in Figs.~\ref{fig:figure_11}b and \ref{fig:figure_11}c.      

The resonant state expansion represents a generalization of the Brillouin-Wigner perturbation theory for non-Hermitian systems~\cite{brillouin1932problemes}. Therefore, numerical accuracy is determined by the size of the basis set $N$. We choose the basis in such a way that for a given orbital number $l$, azimuthal number $n$ and parity we select all resonant states with frequencies lying inside the circle $|\omega R / c|<10$, where $R$ is the radius of the sphere that describes the cylinder.  We consider $l<80$ which results in $N=1035$ that is enough to achieve $99.9\%$ accuracy for calculation of real part of frequencies. Since the perturbation $V_{\alpha\beta}$ conserves the axial symmetry and mirror ($z\rightarrow -z$) symmetry, we study problem for each azimuthal index $n$ and each parity independently.

The dependence of the complex spectrum of eigenmodes with azimuthal indices $n=0,\pm 1$, which are even with respect to up-down reflection symmetry, vs. the cylinder aspect ratio $r/l$ is shown in Fig.~\ref{fig:figure_6} by dotted lines. Dot sizes are proportional to the $Q$ factor. We can clearly observe multiple avoided resonance crossings between modes with the same azimuthal number. In vicinity of an avoided crossing point the $Q$ factor of one of coupled modes dramatically increases, which confirms the results of SCS calculations (Figs.~\ref{fig:figure_1}c and \ref{fig:figure_2}c). 

In the high-frequency region, the behavior of interaction between modes becomes more complicated, e.g. $Q$ factors of some of $n=\pm 1$ modes remain high in a broad range of parameters $x$ and $r/l$, as shown in Fig.~\ref{fig:figure_6}. We explain this phenomenon as strong coupling between three and more eigenmodes with complex spectrum. This broadband high-$Q$ regime will be the subject of our further investigations. In this article we restrict our studies to the mechanism of strong coupling between two eigenmodes because it illustrates the basic peculiarities of spectrum of subwavelength dielectric resonators.

%\bibliography{references}

\end{document}